# Fermi-Level Pinning, Charge Transfer, and Relaxation of Spin-Momentum Locking at Metal Contacts to Topological Insulators

*Catalin D. Spataru\* and François Léonard*

Sandia National Laboratories, Livermore, CA 94551, USA.

ABSTRACT

Topological insulators are of interest for many applications in electronics and optoelectronics, but harnessing their unique properties requires detailed understanding and control of charge injection at electrical contacts. Here we present large-scale *ab initio* calculations of the electronic properties of Au, Ni, Pt, Pd, and graphene contacts to $Bi_2Se_3$. We show that regardless of the metal, the Fermi level is located in the conduction band, leading to n-type Ohmic contact to the first quintuplet. Furthermore, we find strong charge transfer and band-bending in the first few quintuplets, with no Schottky barrier for charge injection even when the topoplogical insulator is undoped. Our calculations indicate that Au and graphene leave the spin-momentum locking mostly unaltered, but on the other hand, Ni, Pd, and Pt strongly hybridize with $Bi_2Se_3$ and relax spin-momentum locking. Our results indicate that judicious choice of the contact metal is essential to reveal the unique surface features of topological insulators.

Topological insulators (TI) exhibit exciting novel properties such as metallic surface states with spin-momentum locking[1]. A number of materials have been predicted[2,3] and observed[4,5] to have the signatures of TIs, holding great promise toward realization of novel applications in spintronics[6] or optoelectronics[7]. In most practical devices, TIs need to be contacted with a metal for electron injection[8,9], as illustrated in Fig. 1a. In this case, important questions can be raised regarding the factors that govern charge or spin transport (Fig. 1b): Are there tunnel or Schottky barriers for electron or hole injection? How important is surface versus bulk injection? Is spin-momentum locking maintained at such interfaces?

In this manuscript, we address these questions by performing large-scale *ab initio* calculations for Au, Pd, Pt, Ni, and graphene contacts to $Bi_2Se_3$. We show that regardless of the metal, the Fermi level is located in the conduction band, leading to n-type Ohmic contact to the first quintuplet. Furthermore, we find strong charge transfer and band-bending in the first few quintuplets, with no Schottky barrier for charge injection even when the topoplogical insulator is undoped. Our calculations indicate that Au and graphene couple relatively weakly to $Bi_2Se_3$, leaving the spin-momentum locking mostly unaltered. On the other hand, Ni, Pd and Pt strongly hybridize with $Bi_2Se_3$ and relax spin-momentum locking. The results indicate that judicious choice of the contact metal is essential to reveal their unique surface features.

Our *ab initio* approach is based on Density Functional Theory (DFT)[10]. We use the generalized gradient approximation (GGA) together with Projector Augmented Wave (PAW) pseudopotentials[11,12]. Spin-orbit coupling (SOC), as implemented in the VASP code[13], is included in the calculations, as well as Van der Waals interactions via semi-empirical corrections using Grimme's method [14,15]. Calculations were performed in a hexagonal supercell geometry[16] with lateral dimensions commensurate with unstrained bulk $Bi_2Se_3$ (with lattice constant of 4.14 Angstroms). We sampled the surface supercell Brillouin zone with grids at least as large as the equivalent 10x10 Monkhorst-Pack mesh per $Bi_2Se_3$ hexagonal cell. The size of the supercell in the case of the $Bi_2Se_3$-Au(111) interface was chosen based on experimental input: STM studies[17] show that a 3 quintuplet $Bi_2Se_3$ layer film on Au(111) forms a Moire pattern with periodicity of ~2.05 nm, suggesting that 5 Bi (or Se) atoms match 7 Au atoms. Thus we chose a supercell of 5x5 $Bi_2Se_3$ matched by 7x7 Au(111). This leads to the metal Au layers being strained laterally by about 2% w.r.t the bulk case. In the other cases, we chose the following lateral supercell sizes: 3x3 $Bi_2Se_3$ matched by 5x5 Ni(111) and 2x2 $Bi_2Se_3$ matched by 3x3 Pd or Pt(111); in these cases the metal layers are strained laterally by less than 1% from their bulk counterpart. A √3x√3 graphene supercell (rotated 30°) was matched by 1x1 $Bi_2Se_3$ after a compressive strain of ~ 3%. (We also considered a second configuration consisting of 3x3 $Bi_2Se_3$ matched by 5x5 unrotated graphene and found similar results as the ones reported here.) Figure 1c shows the general supercell geometry along the direction **z** perpendicular to the interface: in all cases (except graphene) we considered 6 metal layers and 6 $Bi_2Se_3$ quintuplet (QL) layers. The number of atoms in the supercell is 1044 for Au, 420 for Ni, 174 for Pd and Pt, and 36 for graphene.

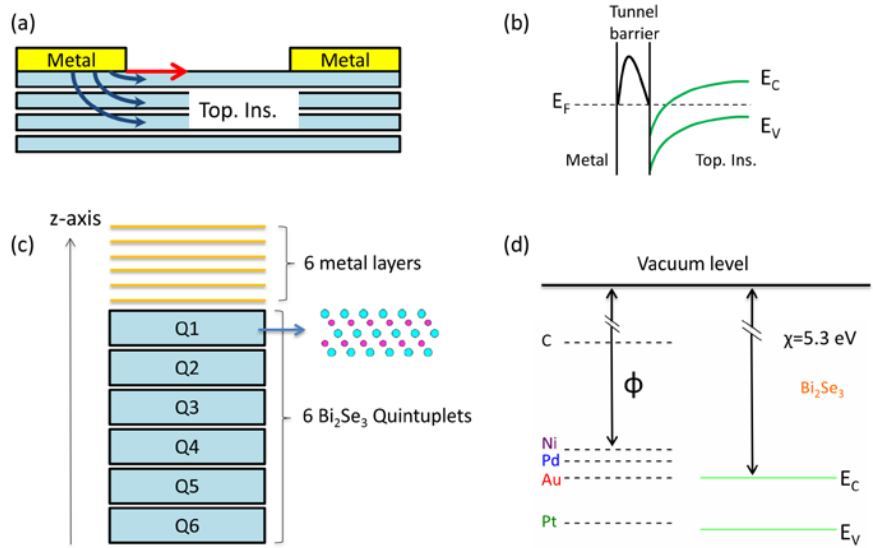

**Figure 1.** (a) Illustration of charge injection between metal contacts and a topological insulator material, with the red (blue) arrow indicating injection in the surface (bulk) states. (b) Sketch to illustrate the important equilibrium properties of metal contacts: the band alignment between the metal Fermi level $E_F$ and the valence $E_V$ and conduction band $E_C$ edges, and the presence of a tunnel barrier. (c) The supercell used in the *ab initio* calculations. (d) Bare alignment between the metal Fermi level and the $Bi_2Se_3$ bands. $\Phi$ is the metal workfunction, $\chi$ is the electron affinity, and a bandgap of 0.4 eV was used.

The equilibrium structure of the system was obtained by relaxing all the forces to better than 10 meV/Angstrom, except for the atoms in the bottom 3 metal layers which were kept fixed (having in mind the situation of a thin TI film in contact with a semi-infinite metal). Due to the large computational demand, relaxation was done without SOC, since it has been shown that SOC has only a minor effect on the atomic structure of a few QLs of $Bi_2Se_3$ [18]; in addition we also verified in the case of $Bi_2Se_3$@Ni(111) that SOC has a negligible effect on the interface structure. Upon structural relaxation, we find an inter-QL distance of about 2.6 Angstrom, while the metal-TI separation $d$ was found to be 3.3, 2.5, 2.2, 2.1 and 2.0 Angstrom for the graphene, Au, Pt, Pd and Ni interfaces, respectively. Relaxed atomic configurations can be found in Fig. S1.

In the simplest picture of the electronic properties of the metal/TI interface, one considers the case of large separation $d$, such that there is negligible overlap between the Ti and metal electron wavefunctions. In that case the alignment between the metal Fermi level and the TI bands is shown in Fig. 1d (before any infinitesimal charge transfer); given the $Bi_2Se_3$ electron affinity of about 5.3 eV, as well as the bulk bandgap of about 0.4 eV, one would expect that Au, Pd, Ni, and graphene would give rise to n-type doping of the TI slab, but that Pt(111) would result in p-type doping. However, our *ab initio* calculations show that for all 5 contacts the metal Fermi level is located in the conduction band of the TI at the surface.

We calculated the band bending potential for each interface as the average electrostatic potential difference between the composite $Bi_2Se_3$@metal system and the isolated $Bi_2Se_3$ thin film, where the Fermi level of the latter was aligned to the metal Fermi level. Fig. 2a shows the calculated band bending potential as a function of the distance from the interface. One can see that near the metal the TI bands can be shifted by as much as 1 eV, thus locating the Fermi level several hundred of meVs above the conduction band edge. The effect is particularly striking for Pt since the Fermi-level pinning gives a n-type contact as opposed to the p-type contact predicted by Fig. 1d. The origin of the pinning can be traced to the large electronegativity of the terminating Se layer. Indeed, $Bi_2Se_3$ in contact with vacuum does not have an accumulated surface, and thus the pinning is not a property of the surface bandstructure. Furthermore, an artificial system with a Bi terminated surface in contact with a metal shows bend-bending of opposite sign[16], in agreement with the small electronegativity of Bi. In the case of graphene, its small workfunction is compensated by the large separation distance and the small density of states, and the Fermi level ends up closer to the conduction band edge than the bareband alignment of Fig. 1d would predict.

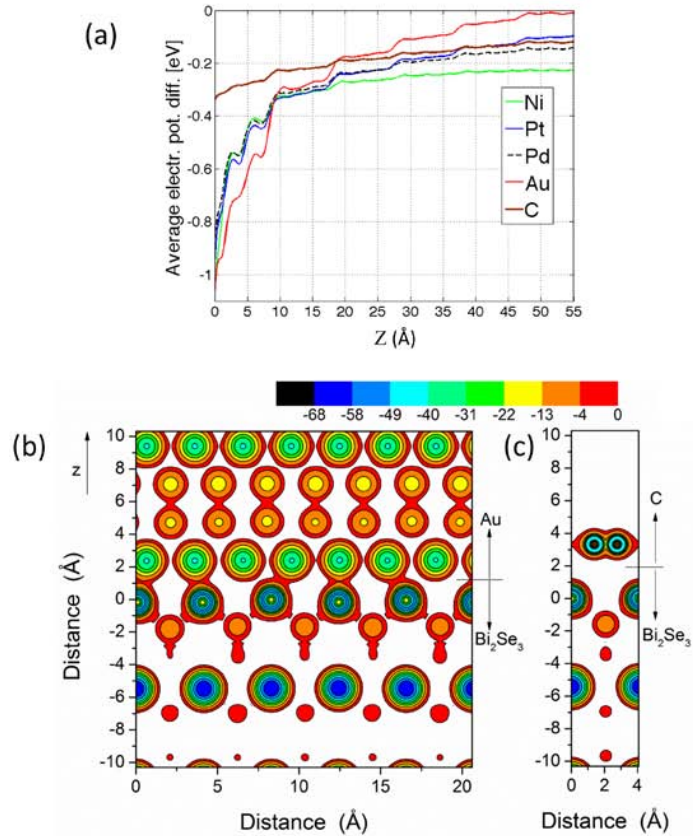

**Figure 2.** (a) Average electrostatic potential difference as a function of distance from the metal interface for Ni, Pt, Pd, Au and graphene. The metal/ $Bi_2Se_3$ interface is at $Z \approx 0$ while the vacuum/ $Bi_2Se_3$ interface is at $Z \approx -5.5$ nm. (b) Color plot of the electrostatic potential (in units of eV) for a cross-sectional plane perpendicular to the interface between Au and $Bi_2Se_3$. Channels across the interface where the potential is below the Fermi level ($E_F = 0$) are present,

implying that there is no tunnel barrier for transport. Similar results are obtained for Au, Ni, and Pt. (c) Similar plot as in (b) for the interface between graphene and $Bi_2Se_3$.

Figure 2b shows a cross-section of the electrostatic potential perpendicular to the Au/TI interface. As can be seen from this figure, there are connecting paths below the Fermi level energy ($E_F = 0$); similar results are found for Ni, Pd, and Pt, implying that for these four metals there is no tunnel barrier for charge injection. In the case of graphene however, because of the large separation and weak coupling no such connecting paths are found, and there are tunnel barriers at the interface (Fig. 2c) that could increase the contact resistance.

Having established the basic properties of the contacts, we now turn to a more in-depth analysis, in particular of hybridization effects and spin-momentum locking. We first consider the case of graphene contacts since this represents a weakly interacting system. We calculate the spectral function projected on quintuplet $j$ using the definition $A_j(\vec{k}, E) = \sum_{n,i}^{i \in QL_j} w_{n\vec{k}}^i \delta(E - \varepsilon_{n\vec{k}})$, where $w_{n\vec{k}}^i$ is the site-projected character of the wavefunction of an electron characterized by band index n and crystal momentum $\vec{k}$ \cite [13], with i the index of the atomic site. In practice the $\delta$-function centered on the electron energy $\varepsilon_{n\vec{k}}$ is broadened by a Lorentzian function of width equal to 10 meV. Figure 3a shows the spectral function projected on the first QL when $Bi_2Se_3$ is in contact with vacuum. The Dirac cone, the TI surface states, as well traces of valence and conduction states located deeper into the TI film are clearly seen in the picture. When in contact with graphene (Fig. 3b), charge transfer leads to the TI Dirac point being shifted down while the Dirac point of graphene shifts up in energy. In addition, traces of the graphene Dirac cone can be distinguished, and one can see that when graphene and TI bands overlap both in energy and momentum, a relatively weak hybridization between the two takes place. Because the band bending potential shifts down the TI surface Dirac point more than the original bulk TI valence bands (deeper into the TI film the band bending potential is smaller – see Fig. 2a) a hybridization between these states can also be noticed. We find that the band dispersion of the graphene states near its Dirac point is essentially unaffected by the TI proximity due to the absence of graphene sublattice symmetry breaking (in the optimized interface geometry –see Fig. S1- the Se atoms avoid being underneath a carbon atom), in contrast to recent model calculations [19] of similar TI-graphene commensurate stacking.

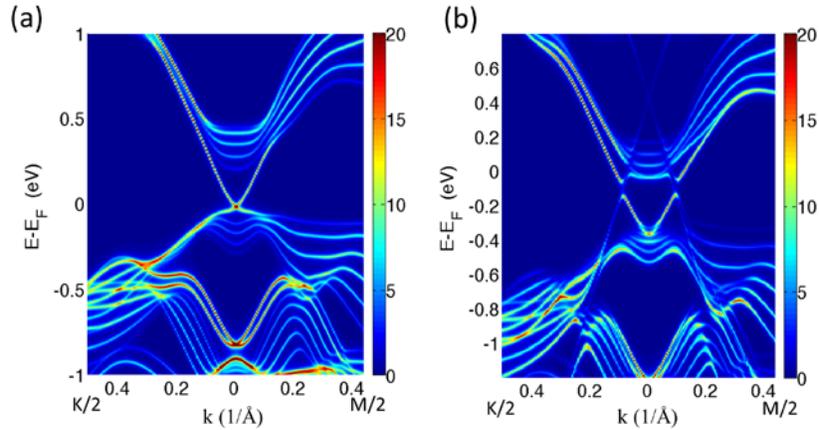

**Figure 3.** (a) Spectral function (in units of 1/eV) projected on the first $Bi_2Se_3$ quintuplet when in contact with vacuum. (b) Spectral function projected on the first $Bi_2Se_3$ quintuplet when in contact with graphene.

When contacted by Au (Fig. 4a), one can still identify the Dirac point, but it is shifted down by about 0.8 eV as expected from the band bending potential for this interface. One can also clearly see the Dirac cone states, which we found to be mostly localized in the first QL. The projected density of states (DOS) for the Au(111) surface shows only a small DOS near the Fermi level, with the large DOS due to d-like states more than 1 eV below the Fermi level (Fig. 4a). This explains why there is relatively little hybridization between TI surface states and Au states.

The situation is different for the case of Pd. From the spectral function projected on the first QL (Fig. 4b) one can see that the Dirac point is hard to identify, with the original Dirac cone states being strongly hybridized with the metal states. Indeed, the lifetime of the resonant states is inversely proportional to the coupling between the TI surface states and the bulk metal ones, as well as to the DOS of the metal bands[20]. The DOS of the Pd(111) surface is large near the Fermi level due to the existence of many flat, d-like metal states in this energy range. We also note the large peak 0.5 eV below the Fermi level, which explains why the lifetime of states corresponding to the upper Dirac cone is much smaller than in the Au interface case, resulting in the blurred resonances. These resonances correspond to states that are delocalized mostly over the metal region. Similar results are obtained for Pt, except that the somewhat smaller DOS leads to slightly sharper resonances below the Fermi level (see Fig. 4c).

We next consider the case of the interface between $Bi_2Se_3$ and the magnetic Ni(111) surface. As seen in Fig. 4d, the spectral function projected on the first QL shows blurred resonances in the energy range where one would expect the Dirac point to be located (based on the band bending potential), and one cannot identify the Dirac cone. This can be explained by the hybridization between TI surface states and metal bands, as suggested by the DOS of the Ni(111) surface. Indeed, one sees in Fig. 4d that the total (majority plus minority spins) DOS shows a minimum in the same energy range (around 0.15 eV below Fermi level) where the blurred resonances are seen. The largest DOS peak takes place around 0.6 eV below the Fermi level (due to the spike of the majority spin DOS), close to the expected energy of the Dirac point based on the bare band alignment of Fig. 1d. (We also simulated a non-magnetic Ni(111) surface, and found the same strong hybridization.)

For all studied metal interfaces electron injection can take place through at least the first two QLs, based on the band-bending (Fig. 2a) and the projected spectral function of the conduction bands in TI that shows no Schottky barrier for electron injection in the second QL[16]. This implies that even for undoped $Bi_2Se_3$ charge injection to the bulk should be efficient, and that p-type counter-doping near the interface may be necessary to favor surface charge injection. (We looked for tunnel barriers between the QLs and found them to be small and narrow.)

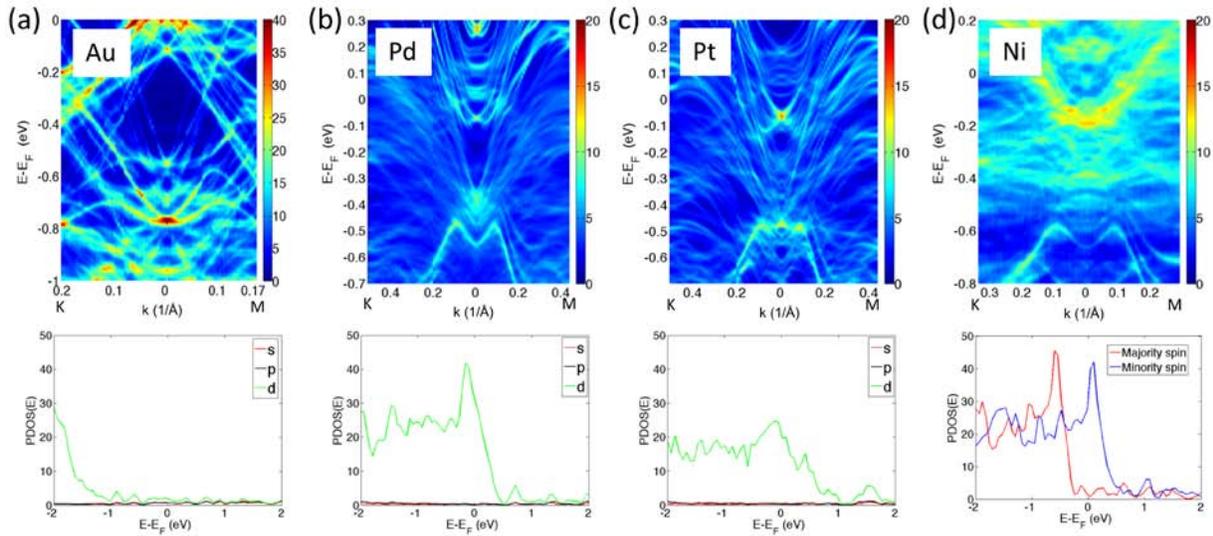

**Figure 4.** The top panels show the spectral function (in units of 1/eV) projected on the first $Bi_2Se_3$ quintuplet when in contact with different metals, and the bottom panels are the density of states of isolated metal slabs with the same normalization used in all 4 cases.

The strength of hydridization between the metal and the TI also has a significant impact on spin-momentum locking. Figure 5 shows again the projected spectral function, but this time only the contribution of states with well-defined spin-momentum locking is retained, namely $\hat{S}_{p_z} \cdot (\hat{k} \times \hat{z}) \approx \pm 1$ where $\hat{S}_{p_z}$ is the direction of the expectation value of the in-plane spin operator for the $p_z$ orbitals [21,22] and $\hat{k}$ is the crystal momentum direction. One can see that for Au, states near the Fermi level have well-defined conventional helicity ($\hat{S}_{p_z} \perp \hat{k}$) with the upper and lower Dirac cones showing states with opposite spin helicity. In the graphene case, the lower Dirac cone is less visible as the states are being pushed toward the second QL by the hybridization with the bulk states [23]. In contrast, for Pd, Pt, and Ni contacts the spin-momentum correlation is relaxed, with little trace of states with spin helicity perpendicular to the crystal momentum. Only when allowing contribution from states with less defined spin-momentum locking, does one start to see a more appreciable trace in the projected bandstructure[16] obtained by retaining states with spin-momentum angles as large as 45 degrees ($\left| \hat{S}_{p_z} \cdot (\hat{k} \times \hat{z}) \right| > 0.7$).

The relaxation of spin-momentum locking is surprising at first in light of the common view that TI surface states are not sensitive to non-magnetic perturbations[24-26], which has been observed in the context of TI interacting with dopant adsorbates [27-29]. However, strong TI-metal interaction can lead to TI surface states being exiled from the metal bandwidth and replaced with resonances[20]. In conjunction with the fact that TI surface states are topologically protected against backscattering but not against scattering at other angles [30], it follows that strong enough perturbations (not necessarily magnetic) can lead to resonances with relaxed spin-momentum locking as our *ab initio* calculations indicate. We also note that our calculations focused on the in-plane spin texture as obtained for the $p_z$ orbitals, but that additional contributions from out-of-plane spin orientations and/or other orbitals should also be considered in the future.

To conclude, we find that for Au and graphene interfaces spin-momentum locking is mostly maintained for the TI surface states. These interfaces provide the most promising metal contacts (of the 5 studied) for spin-polarized transport applications. The Pd and Pt interfaces provide a stronger TI-metal coupling resulting in delocalization of the TI surface states over the metal region and to poor spin-momentum locking. The Ni interface provides the strongest TI-metal coupling, and the TI surface states cannot be distinguished anymore near the metal interface. This interface gives the largest charge transfer from metal to the TI, and may not be appropriate for spin-polarized transport. Taken together, our results suggest that decoupling the metal from the TI may be the most promising route to favor surface injection while maintaining spin-momentum locking. This could be accomplished, for example, by introducing an ultrathin tunnel layer between the metal and the TI.

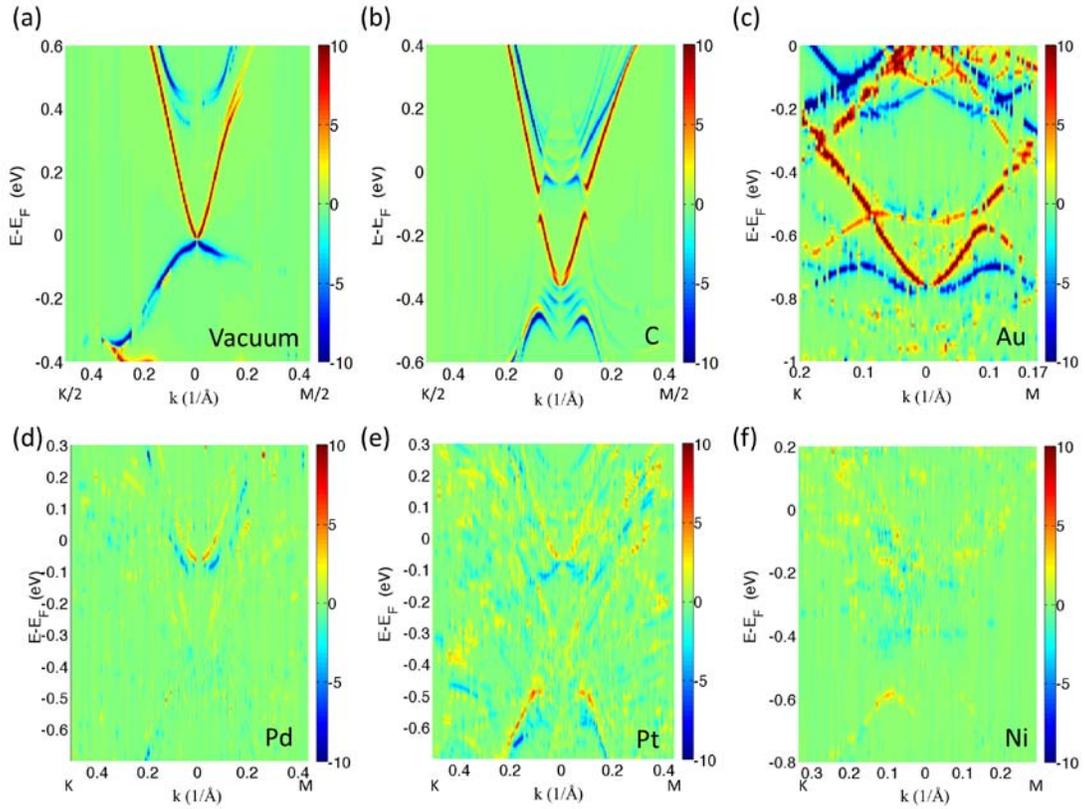

**Figure 5.** Spectral function (in units of 1/eV) projected on the first quintuplet and filtered for spin-momentum locking in the case of (a) vacuum, (b) graphene, (c) Au, (d) Pd, (e) Pt, and (f) Ni.

AUTHOR INFORMATION

**Corresponding Author**

*Email (C.D.S.): cdspata@sandia.gov.

ACKNOWLEDGMENT

We thank Peter Sharma, Michael Shaughnessy and Thomas Beechem for useful discussions. This work was supported by the Electromagnetic Materials program of the Office of Naval Research under agreement #N0001413IP20091 and the Laboratory Directed Research and

# Fermi-Level Pinning, Charge Transfer, and Relaxation of Spin-Momentum Locking at Metal Contacts to Topological Insulators

*Catalin D. Spataru and François Léonard*

Sandia National Laboratories, Livermore, CA 94551, USA.

**Supplementary Material**

1. Atomic structures

As mentioned in the main text, the atomic structures of the metal/$Bi_2S_3$ interfaces were relaxed to obtain the minimum energy configurations. Figure S1 shows the atomic configurations for the 5 types of contacts. Only the Se (yellow) and metal (red) atoms adjacent to the interface are shown. One notes the appreciable reconstruction of the TI surface in the Pd, Pt and Ni cases.

2. Impact of termination on band-bending

To test the idea that electronegativity is responsible for the observed Fermi pinning in the conduction band, we artificially created a Bi-terminated surface by removing the Se layers at the metal and vaccum interfaces. The band bending for this surface has the opposite sign compared to the Se-terminated case (Fig. S2), in agreement with the small electronegativity of Bi.

3. Spectral functions for deeper layers

In the main text, the spectral functions projected on the first quintuplet were shown. In Fig. S3, we show the spectral function for the $2^{nd}$ quintuplet. In the cases b)-f), one can see the trace of a projected conduction band below the Fermi level, indicating no Schottky barrier for electron injection in the second QL. Also, in the Au, Pd and Pt case one can clearly see (at energies ~0.1 eV below the Fermi level) the fingerprint of a second Dirac-like point, derived from the original TI conduction bands that are Rashba-split by the electric field associated with the band bending potential.

4. Relaxation of spin-momentum locking for wider angles

In the main text, the relaxation of spin-momentum locking was tested by using a narrow angle filter of 5 degrees around an angle of 90 degrees between the spin and the momentum. Figure S4 shows similar plots as in the main text, but for a wide filter of 45 degrees. The increased filter range reveals some additional spin-momentum correlation, but it still remains weak overall.

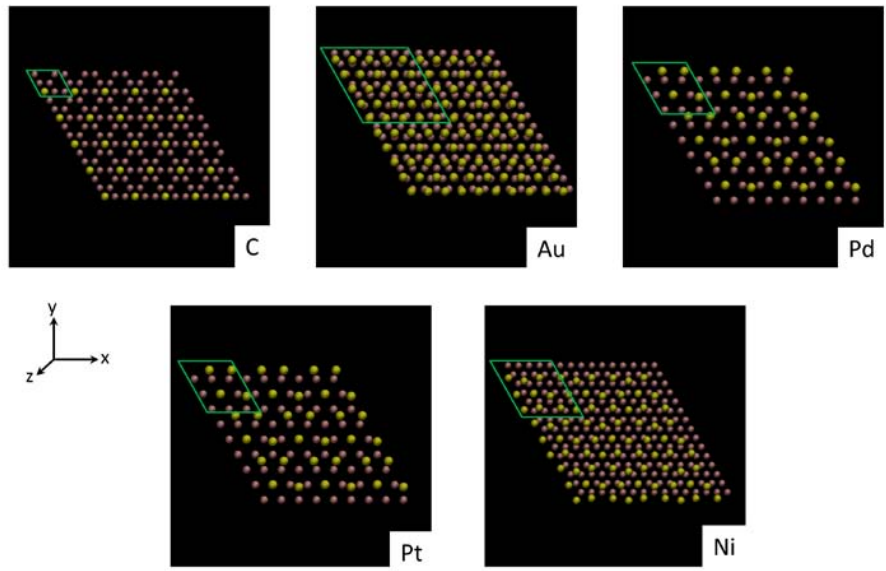

**Figure S1**: Relaxed atomic structures for the 5 different contacts considered. The green boxes denote the unit cell.

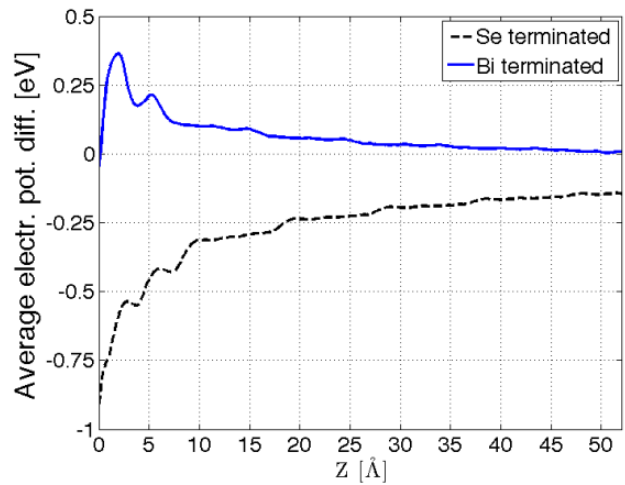

**Figure S2**: Band-bending in the $Bi_2Se_3$ for Pd contacts when the surface termination is artificially changed from Se to Bi (the average metal-TI separation is the same in both cases).

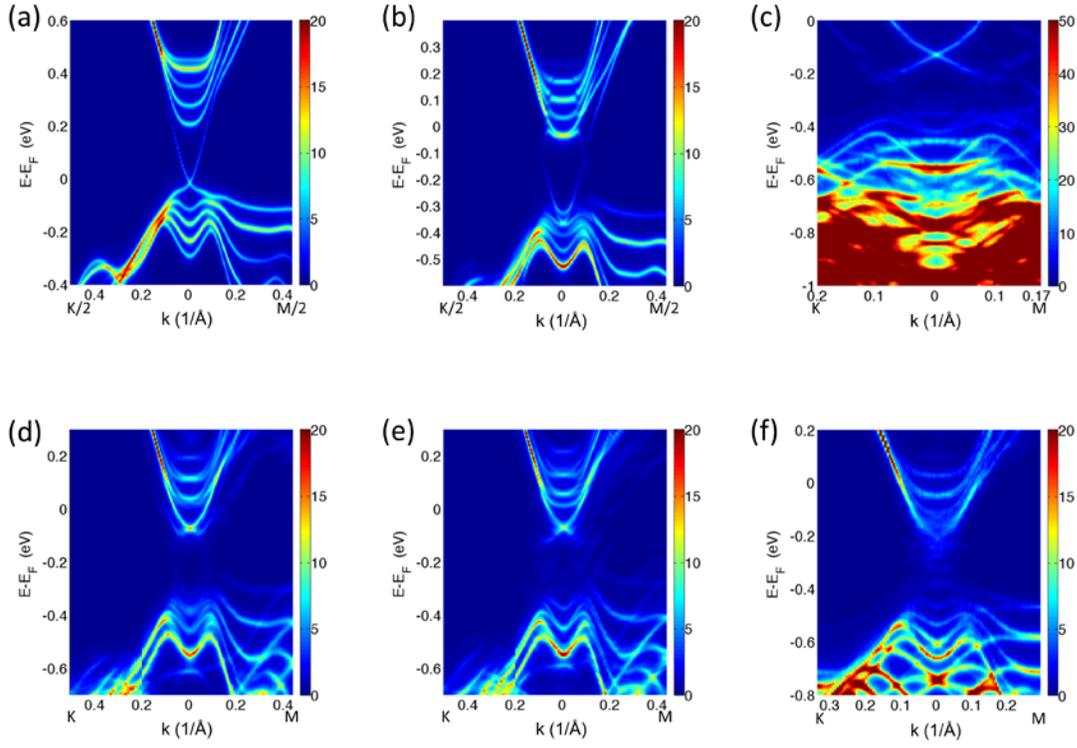

**Figure S3**: Bandstructure projected on the second quintuplet for the 6 cases considered: (a) vacuum, (b) graphene, (c) Au, (d) Pd, (e) Pt, and (f) Ni.

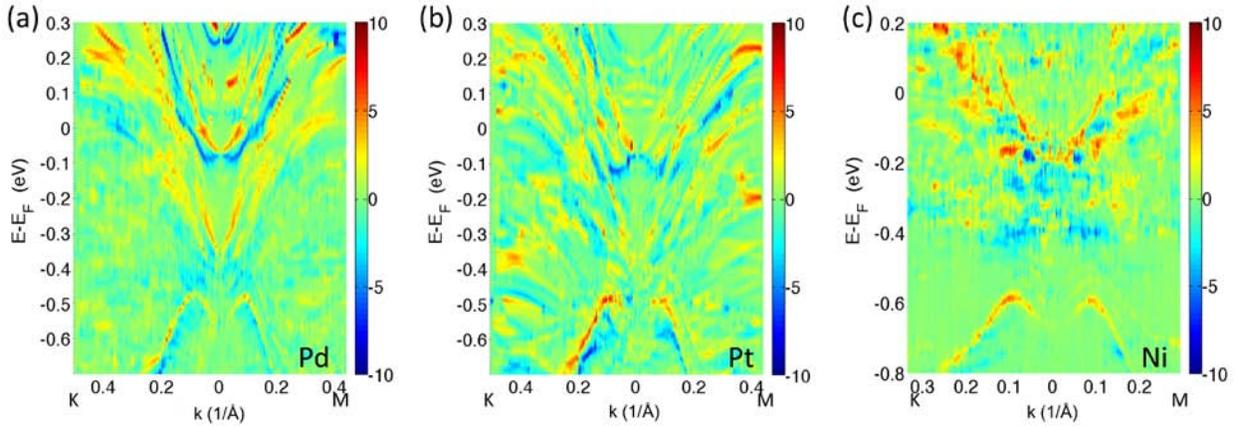

**Figure S4**: Spectral function projected on the first $Bi_2Se_3$ quintuplet and filtered for spin-momentum locking over an angle of 45 degrees.